# Modeling the US-China trade conflict: a utility theory approach

Yuhan Zhang, Cheng Chang

Yuhan Zhang: UC Berkeley, CA, USA zyhberkeley2020@berkeley.edu
Cheng Chang: Mercy College, NY, USA cchang4@mercy.edu

Corresponding author: Cheng Chang
Mercy College Dept. of Mathematics & Computer Sciences
555 Broadway, Dobbs Ferry, NY 10522, USA

## ABSTRACT
This paper models the US-China trade conflict and attempts to analyze the (optimal) strategic choices. In contrast to the existing literature on the topic, we employ the expected utility theory and examine the conflict mathematically. In both perfect information and incomplete information games, we show that expected net gains diminish as the utility of winning increases because of the costs incurred during the struggle. We find that the best response function exists for China but not for the US during the conflict. We argue that the less the US coerces China to change its existing trade practices, the higher the US expected net gains. China's best choice is to maintain the status quo, and any further aggression in its policy and behavior will aggravate the situation.

# 1. Introduction

The trade conflict between the world's two largest economies—the United States (US) and China—initiated in 2018 has captured massive attention among academics and decision-makers. Major strands of literature discuss the direct economic costs of tariffs (Amiti et al. 2019, Itakura 2019) and repercussions on the international trading system (Lawrence 2018). Some scholars examine the causes of the friction and point out the crux of the issue involves not only the balance of payments but also technological competition (Zhang 2018, Chen et al. 2019). A few researchers analyze the issue through the lens of game theory, yet their studies are rather qualitative (Yin and Hamilton 2019, Jiang et al. 2020).

This paper attempts to model the bilateral conflict and understand the US' and China's strategic choices in mathematical terms. It presents an extension to the literature on the expected utility theory. The high-level theoretical framework designed in this paper can be applied not only to the economic conflict but to the technology friction between the two powers.

Models in this paper capture both players' calculations on expected net benefits from fighting and consider two different situations with perfect information and incomplete information. Fighting in this paper does not mean a sequential trade conflict; instead, it refers to economic coercion and backfires. The paper solves inequality constrained optimization and proves the existence or non-existence of the extrema of expected net gains. Three findings are particularly worth noting. First, when each player's utility of winning increases, their expected net gains diminish. Second, the US best response function exists neither in the perfect nor the incomplete information games. Third, the best response for China is to preserve the status quo.

# 2. Models

In the trade conflict, each player has a chance to either win or lose. For the US, winning means China will be forced to change its trade practice and technology policy while losing indicates that, with the US coercion (e.g., impose massive tariffs on imported Chinese goods), China will not change its existing policies (aka the status quo) or its policies become even more aggressive (e.g., depreciate its currency and enlarge its current account surplus or slash imports from the US). On the flip side, winning and losing outcomes for China are the exact opposite of those of the US.

The probabilities of winning and losing are measured in terms of relative capabilities (Cao 2013). In our context, the capacities of the US and China are substantial, constant values, denoted as $C_i$, i = {US, CN}. US capability is more significant than China's.[1] Hence, the probability that the US wins can be written as $\alpha = \frac{C_{US}}{C_{US}+C_{CN}} > 0.5,$ and the probability that China wins can be expressed as $\beta = \frac{C_{CN}}{C_{US}+C_{CN}} < 0.5.$

Because the US and China are rational players, the objective for each is to maximize their expected net gains of fighting. The net gains of player i consist of three components—the utility of fighting, cost, and the utility of not fighting (status quo), which will be defined in the sections below.

## 2.1. Perfect information

Utilities of winning and losing are denoted by $U_W^i$ and $U_L^i$, respectively, where W represents winning and L means losing. When player i wins, the cardinal utility will be greater than the one for losing, that is, $U_W^i > U_L^i$.

The cost of fighting is a function of the adversary's incentive to retaliate economically and its capacity. Under perfect information, both players have full knowledge of the opponent's incentive. Let $I(U_k^{US})$ denote the incentive function for China to blowback (e.g., add tariffs on the US goods or devalue the Chinese currency), where k = {W, L}. We assume $I$ is an increasing odd function and $I \in (-1,1)$. Additionally, $I$'s first-order derivative, $I'$, is bounded. Specifically, we assume $\frac{1}{C_{CN}} < I' < 1$. Notice that $\frac{1}{C_{CN}}$ is a very small positive value. After scaled by China's capacity, the cost to the US is $C_{CN} I(U_k^{US})$. Similarly, define $\psi(U_k^{CN})$ as the US incentive function to fight back (e.g., add more tariffs on Chinese goods to threaten China to accede to the US demand). $\psi$ is also an increasing odd function and $\psi \in (-1,1)$. Moreover, $\frac{1}{C_{US}} < \psi' < 1$. China's cost is $C_{US} \psi(U_k^{CN})$.

Lastly, the utility of the status quo is denoted $U_S^i$. The expected net gains are respectively

$$E_{US} = \alpha[U_W^{US} - C_{CN} I(U_W^{US}) - U_S^{US}] + (1 - \alpha)[U_L^{US} - C_{CN} I(U_L^{US}) - U_S^{US}] \qquad (1)$$

$$E_{CN} = \beta[U_W^{CN} - C_{US} \psi(U_W^{CN}) - U_S^{CN}] + (1 - \beta)[U_L^{CN} - C_{US} \psi(U_L^{CN}) - U_S^{CN}] \qquad (2)$$

As mentioned earlier, if the US cannot alter China's policies and behaviors, the state is defined as "lose". If the US can change China's, the state is defined as a "win". Also, the US utility of winning is China's payoff of losing, and the US payoff of losing is China's utility of winning, so we have $U_W^{CN} = U_S^{CN} = -U_L^{US} = -U_S^{US}$ and $U_W^{US} = -U_L^{CN}$.

Hence, Eq. (1) and Eq. (2) are simplified to

$$E_{US} = \alpha[U_W^{US} - C_{CN} I(U_W^{US}) - U_S^{US}] - (1 - \alpha) C_{CN} I(U_S^{US}) \qquad (3)$$

$$E_{CN} = -\beta C_{US} \psi(U_W^{CN}) + (1 - \beta)[U_L^{CN} - C_{US} \psi(U_L^{CN}) - U_W^{CN}] \qquad (4)$$

## 2.2. Incomplete information

To make our analysis more engrossing and realistic, we consider uncertainty in the incentive functions. For instance, factors such as internal economic problems might impact the commitment to the game effort. The opponent only knows the probability distribution of those factors.

Here we define the incentive function for China to fight back as $I(U_k^{US}, z)$ where $z \in [0,1]$ is a random variable (e.g., the impact of a pandemic on the domestic economy) that is independent of $U_k^{US}$. $f(z)$ is the probability density function. Likewise, the incentive function for the US is $\psi(U_k^{CN}, \varepsilon)$ where random variable $\varepsilon \in [0,1]$ and $g(\varepsilon)$ is the probability density function.

We adopt the same concept of expected net gains in 2.1. We have

$$E_{US} = \alpha[U_W^{US} - C_{CN} \int I(U_W^{US},z)f(z)dz - U_S^{US}] - (1-\alpha) C_{CN} \int I(U_S^{US},z)f(z)dz \qquad (5)$$

$$E_{CN} = -\beta C_{US} \int \psi(U_W^{CN},\varepsilon)g(\varepsilon)d\varepsilon + (1-\beta)[U_L^{CN} - C_{US} \int \psi(U_L^{CN},\varepsilon)g(\varepsilon)d\varepsilon - U_W^{CN}] \qquad (6)$$

Notice that incentive functions under incomplete information are obtained from the original incentive functions by shifting, depending upon $z$ and $\varepsilon$. Specifically, $I(U_k^{US},z) = I(U_k^{US},z) - M(z)$ and $\psi(U_k^{CN},\varepsilon) = \psi(U_k^{CN}) - N(\varepsilon)$ where $M$ and $N$ are non-negative, strictly increasing functions.

## 3. Analytical results

We explore our models and calculate the extrema of $E_i$ through mathematical analysis. The following propositions state our main analytical results.

**Proposition 1**
Consider $E_{US}$ in Eq. (3) subject to
i. $0 < U_W^{US} \leq \underline{c}$
ii. $-\overline{c} \leq U_S^{US} \leq -\underline{c}$

When China's policies are not changed (the status quo remains), $U_W^{CN} = -U_S^{US} = \underline{c}$. When China's behaviors become more undesirable to the US, $U_W^{CN}$ increases and can reach $\overline{c}$. $\underline{c}$ and $\overline{c}$ are positive constants and $\overline{c}$ is much greater than $\underline{c}$.

The minimum achieves when $U_W^{US} = -U_S^{US} = \underline{c}$. The minimum value of $E_{US}$ is $2\alpha\underline{c} + (1-2\alpha)C_{CN} I(\underline{c}) < 0$. The maximum of $E_{US}$ does not exist. Neither does the best response function for the US.

**Proof**
Notice that the set defined by the three conditions, denoted D, is a pre-compact set in $R^2$. Hence, the closure of D, say $\overline{D} \subseteq R^2$, is a compact subset.

Actually, $\overline{D} = \{0 \leq U_W^{US} \leq \underline{c}\} \cap \{-\overline{c} \leq U_S^{US} \leq -\underline{c}\}$. Since $E_{US}$ is continuous in $\overline{D}$, so both maximum and minimum exist. We will first find the extreme values of $E_{US}$ in $\overline{D}$, then show that the maximum occurs in $\overline{D} - D$ and the minimum occurs inside D.

We only consider $E_{US}$ as a function of $U_W^{US}$ as $U_S^{US}$ is determined by China in the game. The partial derivative is $\frac{\partial E_{US}}{\partial U_W^{US}} = \alpha[1 - C_{CN} I'(U_W^{US})]$.

By the assumption of $\frac{1}{C_{CN}} < I' < 1$, $\frac{\partial E_{US}}{\partial U_W^{US}} < 0$. Therefore, $E_{US}$ achieves minimum in D when $U_W^{US} = -U_S^{US} = \underline{c}$. The minimum value is $2\alpha\underline{c} + (1-2\alpha)C_{CN} I(\underline{c})$. Because we consider $C_{CN}$ is much larger than $\underline{c}$, so the minimum value is less than 0. On the other hand, $E_{US}$ has a maximum in $\overline{D} - D$ when $U_W^{US} = 0$, so maximum for $E_{US}$ does not exist in D.

**Proposition 2**

Consider $E_{CN}$ in Eq. (4) subject to

i. $\underline{c} \leq U_W^{CN} \leq \bar{c}$
ii. $-\underline{c} \leq U_L^{CN} < 0$

There exists a maximum for $E_{CN}$ when $U_W^{CN} = \underline{c} = -U_L^{CN}$, that is, China maintains the status quo, and policies are unchanged. At this point, China's strategy produces the most favorable outcome. The minimum exists when $U_W^{CN} = \bar{c}$, that is, China behaves to a certain level that is worse than the status quo despite the US' coercion. The maximum value is

$(1 - 2\beta)C_{US}\psi(\underline{c}) - 2(1 - \beta)\underline{c} > 0$, and the minimum value is
$C_{US}[-\beta\psi(\bar{c}) - (1 - \beta)\psi(U_L^{CN})] + (1 - \beta)[U_L^{CN} - \bar{c}]$ which is likely less than 0.

**Proof**

Consider $E_{CN}$ in the compact set: $\{\underline{c} \leq U_W^{CN} \leq \bar{c}\} \cap \{-\underline{c} \leq U_L^{CN} \leq 0\}$.

We only consider $E_{CN}$ as function of $U_W^{CN}$ as $U_L^{CN}$ is determined by the US. The partial derivative is $\frac{\partial E_{CN}}{\partial U_W^{CN}} = -\beta C_{US}\psi'(U_W^{CN}) - (1 - \beta)$.

Hence, $\frac{\partial E_{CN}}{\partial U_W^{CN}} < 0$. Given conditions i. and ii., the maximum occurs when $U_W^{CN} = \underline{c} = -U_L^{CN}$, the best response function exists, and the minimum occurs when $U_W^{CN} = \bar{c}$. Because we consider $\bar{c}$ is much larger than $\underline{c}$ and less than $C_{US}$, the maximum value is greater than 0, and the minimum value is likely less than 0.

**Proposition 3**

Under incomplete information, assume $E_{US}$ in Eq. (5) is subject to the same conditions in Proposition 1.

As $U_W^{US}$ increases, $E_{US}$ decreases. The minimum of $E_{US}$ achieves when $U_W^{US} = -U_S^{US} = \underline{c}$. The minimum value is $2\alpha\underline{c} + C_{CN}[\int M(z)f(z)dz + (1 - 2\alpha)I(\underline{c})]$. The maximum does not exist. The best response function cannot exist for the US.

**Proof**

Based on Eq. (5), we have the partial derivative $\frac{\partial E_{US}}{\partial U_W^{US}} = \alpha[1 - C_{CN}\int \frac{\partial I(U_W^{US},z)}{\partial U_W^{US}} f(z)dz]$.
By the assumption that $I(U_W^{US}, z) = I(U_W^{US}) - M(z)$ and $\frac{1}{C_{CN}} < I' < 1$,
$\frac{\partial E_{US}}{\partial U_W^{US}} = \alpha[1 - C_{CN}I'(U_W^{US})] < 0$

Thus, $E_{US}$ has a minimum at $U_W^{US} = -U_S^{US} = \underline{c}$. The minimum value is $2\alpha\underline{c} + C_{CN}[\int M(z)f(z)dz + (1 - 2\alpha)I(\underline{c})]$. Notice that when $M(z)$ is a small number, the minimum value is likely less than 0. The maximum does not exist, as $U_W^{US} \to 0$ but $U_W^{US} \neq 0$.

**Proposition 4**

Consider $E_{CN}$ in Eq. (6), subject to the same conditions in Proposition 2.

There exists a maximum for $E_{CN}$ when $U_W^{CN} = \underline{c} = -U_L^{CN}$, and a minimum when $U_W^{CN} = \bar{c}$. The maximum value is $C_{US}[(1-2\beta)\psi(\underline{c}) + \int N(\varepsilon)g(\varepsilon)d\varepsilon] - 2(1-\beta)\underline{c}$, and the minimum value is $-\beta C_{US}\psi(\bar{c}) + C_{US}\int N(\varepsilon)g(\varepsilon)d\varepsilon + (1-\beta)[U_L^{CN} - C_{US}\psi(U_L^{CN}) - \bar{c}]$. The best response function exists for China.

**Proof**

Based on Eq. (6), we have the partial derivative

$$\frac{\partial E_{CN}}{\partial U_W^{CN}} = -\beta C_{US} \int \frac{\partial \psi(U_W^{CN}, \varepsilon)}{\partial U_W^{CN}} g(\varepsilon)d\varepsilon - (1-\beta)$$

Because $0 < \beta < 0.5$ and the expectation of $\frac{\partial \psi(U_W^{CN}, \varepsilon)}{\partial U_W^{CN}}$ is greater than 0, $\frac{\partial E_{CN}}{\partial U_W^{CN}} < 0$. Given the conditions, the best response function does not exist. $E_{CN}$ reaches the minimum when $U_W^{CN} = \bar{c}$. The minimum value is $-\beta C_{US}\psi(\bar{c}) + C_{US}\int N(\varepsilon)g(\varepsilon)d\varepsilon + (1-\beta)[U_L^{CN} - C_{US}\psi(U_L^{CN}) - \bar{c}]$. The maximum exists with the value of $C_{US}[(1-2\beta)\psi(\underline{c}) + \int N(\varepsilon)g(\varepsilon)d\varepsilon] - 2(1-\beta)\underline{c}$. Notice that when $N(\varepsilon)$ is a small number, the minimum value is likely less than 0.

## 4. Discussion and conclusion

This study examines the protracted US-China trade conflict mathematically. Our approach contributes to the existing literature on the expected utility theory. We construct models that tailor the US-China case. Our high-level theoretical framework can also be applied to examine the bilateral technology flare-up.

Unlike other studies, we focus on the conflict and understand the US and China's strategic choices in mathematical terms. Based on our models, we demonstrate that the utility of winning will reduce the expected net gains of the fighting because of the costs incurred during the struggle. We further prove that the US cannot find a strategy to produce the optimal outcome of the bilateral conflict. Hence, the less the US uses protectionist tariffs to force China to change its trade practices, the higher the expected net gains. Interestingly, China has the best response function if it maintains its existing trade policies. However, we show that any further aggressive policy and behavior, such as enlarge trade surpluses or slash imports from the US, will exasperate the US and aggravate the situation.

This work is exclusively theoretical. Future research could perform a statistical analysis, which may help better understand decision-making dynamics.

## Disclosure statement

The authors reported no potential conflict of interest.

---

[1] World Bank (2019) published GDP ranking, US News (2020) published overall capacity ranking, and the Chinese Academy of Social Sciences (2019) published technology capability ranking. All indicate US capability is greater than China's.